\documentclass[twocolumn]{autart}    

\usepackage[authoryear]{natbib}
\usepackage{graphicx}          

\usepackage{amsmath,amsfonts}
\usepackage{array}

\usepackage{textcomp}
\usepackage{stfloats}
\usepackage{url}
\usepackage{verbatim}
\usepackage{subfigure}
\usepackage{color}
\usepackage{amssymb}
\usepackage{textcomp}
\usepackage{algorithm}  
\usepackage{algpseudocode}  
\usepackage{amsmath}  
  
\usepackage[justification=centering]{caption}
\usepackage{mathtools} 
\usepackage{booktabs}    

\renewcommand\t {^{\mbox{\tiny\sf T}}}

\newtheorem{definition}{Definition}
\newtheorem{lemma}{Lemma}
\newtheorem{problem}{Problem}
\newtheorem{assumption}{Assumption}
\newtheorem{remark}{Remark}
\newtheorem{corollary}{Corollary}
\newcommand{\bproof}{{ \it Proof: }}
\newcommand{\eproof}{\hfill\rule{2mm}{2mm}}

\begin{document}

\begin{frontmatter}

\title{Minimal-time Deadbeat Consensus and  Individual  Disagreement Degree Prediction for  High-order Linear Multi-agent Systems\thanksref{footnoteinfo}} 

\thanks[footnoteinfo]{The material in this paper was not presented at any conference. Corresponding author: Hai-Tao Zhang.}

\author[HUST]{Fu-Long Hu}\ead{fulong@hust.edu.cn},  
\author[HUST]{Hai-Tao Zhang}\ead{zht@mail.hust.edu.cn},    
\author[NPU]{Bowen Xu}\ead{xubowenjason@163.com} ,
\author[HUST]{Zhe Hu}\ead{zhehu@hust.edu.com} , 
\author[UCR]{Wei Ren}\ead{ren@ece.ucr.edu}  

\address[HUST]{School of Artificial Intelligence and Automation, Key Laboratory of Image Processing and Intelligent
	Control, and the State Key Laboratory of Digital Manufacturing Equipment
	and Technology, Huazhong University of Science and Technology, Wuhan,
	430074, China.}  
\address[NPU]{School of Artificial Intelligence, Optics and Electronics, Northwestern Polytechnical University, Xi'an 710072, China}             
\address[UCR]{Department of Electrical and Computer Engineering, University of California, Riverside 92521, USA}


\begin{abstract}                          
  \hspace{0.2 in}In this paper, a Hankel matrix-based fully distributed algorithm is proposed to address a minimal-time deadbeat consensus prediction problem for discrete-time high-order multi-agent systems (MASs). Therein, each agent can  predict  the consensus value with  the minimum number of observable historical outputs of its own.  Accordingly, compared to  most  existing  algorithms only  yielding   asymptotic convergence,  the present method can attain  deadbeat consensus  instead.  Moreover, based on the consensus value prediction, instant individual disagreement degree value of MASs can be calculated in advance as well. Sufficient conditions are derived to guarantee both the minimal-time deadbeat consensus and the instant individual disagreement degree prediction. Finally, both the effectiveness and superiority of the proposed deadbeat consensus algorithm are substantiated by numerical simulations.
\end{abstract}

\end{frontmatter}
\section{Introduction}
\label{sec:introduction}
 \vspace{-0.5em} \hspace{0.2 in}Due to their high efficiency, wide coverage and low cost, a large volume of efforts have been devoted to the distributed control of multi-agent systems (MASs). Therein, consensus situation is a requisite that refers to a certain state (i.e., acceleration, position, voltage) of all the agents reaching an agreement merely by using local interaction \citep{ren2008distributed,olfati2006flocking}. To this end, quite a few consensus algorithms have been proposed over the past few decades \citep{ren2005consensus,fax2004information,dimarogonas2007rendezvous,huang2014overview}, which have been found universal applications to smart grids \citep{yang2013consensus},  social networks  \citep{proskurnikov2015opinion}, multi-sensor networks \citep{sun2017multi}, unmanned systems \citep{liu2019collective}, etc. So far, most of the existing relevant approaches focus on asymptotically attaining consensus by utilizing  local interactions \citep{olfati2004consensus,jadbabaie2003coordination,moreau2005stability,xiao2004fast}, which update each agent's state by some kinds of weighted linear combinations of itself and its neighbors.   Therein, exact convergence to the consensus state could not be achieved within finite steps, which could not fulfill more and more strict  cooperation efficiency  requirements of  modern industrial MASs.

 \vspace{-1em} \hspace{0.2 in}In the past decades,  as a remedy, a  promising kind of scheme has been proposed, which calculates the  exact final consensus value  according to  sequential  individual states  \citep{charalambous2018stop,zhang2008collective,zhang2019ultrafast,manitara2017distributed,sundaram2010distributed}. As a pioneer work, a consensus prediction scheme for first-order linear MASs with time-invariant topology was proposed in \citep{sundaram2007finite}, where each individual computes the consensus value using  finite historical sampling series  stored in its memory. Enlightened by  such representative works,  \citep{charalambous2018stop,zhang2008collective,zhang2019ultrafast,manitara2017distributed,sundaram2010distributed,sundaram2007finite}, some efforts \citep{yuan2012decentralised,yuan2013}  have further reduced the individual consensus prediction cost by substantially decreasing the total historical state series length required to be stored in individual memory. Following this research line, \citep{charalambous2015distributed} proposed a distributed algorithm that calculate  exact average consensus values in a finite period for first-order MASs with time-delays.  

 \vspace{-1em} \hspace{0.2 in}Till date, most of the aforementioned consensus value prediction algorithms focus on first-order MASs and { adopt the final value theorem of $\mathcal{Z}$-transform to calculate the  eventual consensus values, which can not be directly extended to high-order MASs since the limits of their eventual values do not always exist.} Due to the challenges for high-order evolution calculation  merely by local information exchanging,  so far, minimal-time deadbeat  consensus  prediction for  high-order linear MASs has not been touched. However, for both natural biological  swarms/flocks/schools and engineering collective motion systems, it is often required for collective motional systems to take prompt reactions to environmental emergencies,  urgent tasks or  even suddenly-appeared group targets. In such  universal   situations, development of a  deadbeat  consensus  prediction  protocol becomes indispensable for  high-order MASs, which needs to reach exact agreement upon not only  positions and velocities, but also accelerations and  jerks \citep{attanasi2014information,cavagna2015flocking}, with sufficiently short individual memory lengths.  Therefore, consensus protocol design of high-order MASs has attracted more and more attention from relevant researchers \citep{rezaee2015,seo2009consensus,7122276,7984883, hu2019decentralized}. This motivates us to consider an  urgent yet challenging problem, i.e., designing a decentralized protocol to predict  the exact eventual consensus values of high-order linear MASs with directed topological backbones merely according to individual memory as short as possible. 

 \vspace{-1em} \hspace{0.2 in}In brief, the main contribution of this paper is as below. A minimal-time deadbeat consensus prediction (MDCP) protocol is  developed for discrete-time high-order linear MASs. Sufficient conditions are theoretically derived to guarantee  the convergence of   MDCP.  Moreover, rather than just prediction of steady-state consensus values of MASs, the present MDCP could predict the instant states of each  individual of the MASs as well, which  in turn describes the future evolution of  the individual disagreement degree among the MASs more delicately that is desirable for more complicated  real application cases like rendezvous, coverage, flocking and formation control.

 \vspace{-1em} \hspace{0.2 in}The remainder of this paper is organized as follows. In Section~II,  the main problem addressed by the present study is presented together with the preliminaries.  Then,  MDCP is designed in Section~III  and the sufficient conditions are derived to guarantee the convergence of  MDCP. Afterwards, numerical simulations are conducted in Section~IV to verify the effectiveness of the present protocol. Finally,  conclusion is drawn in Section~V.

 \vspace{-1em} \hspace{0.2 in}Throughout this paper, the following symbols will be used. $\mathbb{R}$, $\mathbb{R}^n$, $\mathbb{R}^{n \times m}$, $\mathbb{R}^+$ refer to the sets of real number, $n$-dimensional real vector, $n \times m$ real matrix and positive real number, respectively.  $\mathbb{N}$, $\mathbb{Z}$ and $\mathbb{Z}^+$ indicate  the natural number, integer number and positive integer number sets, respectively. The transpose of  a matrix $\ast$ is denoted by $\ast\t $. Size$(\ast)$ indicates the rows of a square matrix $\ast$, and $\ast^\bot$ denotes the null space of $\ast$.   $(e_n^{i})\t =\left[0, \ldots, 0,1_{i \text {-th }}, 0, \ldots, 0\right] \in \mathbb{R}^{1 \times n}$ is an $n$-dimension row vector with  $i$-th element being 1 and the others 0. The symbol det$(\ast)$  denotes the determinant of a matrix $\ast$. $\otimes$ refers to the Kronecker product. $\textbf{1}_n \in \mathbb{R}^{n}$ and $\textbf{0}_n \in \mathbb{R}^{n}$ are the column vectors with all elements being 1 and 0, respectively. $\textbf{0}$ is a matrix with compatible dimensions and all elements being 0. $I_n \in \mathbb{R}^{n \times n}$ denotes an $n \times n$ identity matrix. The $\mathcal{Z}$-transform of a function $f$ is denoted by  $\mathcal Z(f)$.  The symbol $ \lfloor \ast\rfloor:=\max \{n \in \mathbb{Z} \mid n < \ast\}$ indicates the lower integral function. Euclidean norm of a vector $*$ is denoted by $\left\| *\right\| $.  The symbol $_{i}^{j}*:=\left[  *(i) ,*(i+1),\dots,*(j)\right]  \in \mathbb{R}^{j-i+1},j>i,i,j \in \mathbb{N}$ denotes a row vector, where $*(i)$ refers to the $i$-th element of $*$. $\sup(*)$ denotes the supremum of set $*$. 
 \vspace{-0.5em}
\section{Preliminaries and problem formulation}
 \vspace{-0.5em} \hspace{0.2 in}Consider a fixed directed graph  consisted of $n$ agents given by $\mathcal{G}=\left( \mathcal{V}, \mathcal{E}, \mathcal{A}\right) $, where $\mathcal{V}=\left\lbrace 1,2,\dots ,n\right\rbrace $ is a finite nonempty  node set denoting the agents, $\mathcal{E} \subseteq \mathcal{V} \times \mathcal{V}$ is the edge set. $\mathcal{A}=\left[a_{i j}\right] \in \mathbb{R}^{n \times n}$ represents the weighted adjacency matrix. Here, $a_{ij} > 0$ if $(j, i) \in \mathcal{E}$, and $a_{ij}=0$ otherwise, furthermore, only simple graph is considered. i.e., there is no self-loop. Let $\mathcal{L}:=[\mathcal{L}_{ij}] \in \mathbb{R}^{n \times n}$ denotes the Laplacian matrix, where $\mathcal{L}_{ii} = \sum_{i \neq j}a_{ij}$, $\mathcal{L}_{ij}=-a_{ij}$ for $i \neq j$. If directed graph $\mathcal{G}$ has a directed spanning tree, 0 is a simple eigenvalue of $\mathcal{L}$ corresponding to eigenvector $\textbf{1}_n$, i.e., $\mathcal{L}\textbf{1}_n=0$.

 \vspace{-1em} \hspace{0.2 in}Consider a discrete-time $s$-order linear MAS consisted of $n$ agents with dynamics described by  
\begin{equation}
	\begin{split}
		x_i^{(1)}(k) &=x_i^{(1)}(k-1)+\epsilon x_i^{(2)}(k-1), \\
		x_i^{(2)}(k) &=x_i^{(2)}(k-1)+\epsilon x_i^{(3)}(k-1), \\
		& \vdots \\
		x_i^{(s-1)}(k) &=x_i^{(s-1)}(k-1)+\epsilon x_i^{(s)}(k-1),\\
		x_i^{(s)}(k) &=x_i^{(s)}(k-1)+ u_i(k-1),
	\end{split}
	\label{dynamic}
\end{equation}  
where $x_i^{(s)}(k) \in \mathbb{R}^{m}$ denotes the $s$-order state  of $i$-th agent  at the $k$-th time instant, $i=1,2,\cdots,n$, $s,k \in \mathbb{Z}^+$, and $\epsilon \in \mathbb{R}^+$ represents the sampling time. Without loss of generality,  $m=1$ is considered throughout this paper. The scenarios for $m>1$ can be easily extended with the assistance of the Kronecker product `$\otimes$'.

 \vspace{-1em} \hspace{0.2 in}Conventional control protocol  for high-order linear MASs is given as follows \citep{ren2007high} \vspace{-2em} 
\begin{small}
	\begin{equation}
		u_i(k-1)=\omega\sum_{j=1}^n a_{i j}   p_{s}\left(x_i^{(1)}(k-1)-x_j^{(1)}(k-1)\right)  \label{ui},
	\end{equation}
\end{small}
where $p_{s}(x):=c_{0} x+c_1 x^{(1)}+\cdots+c_{s-1} x^{(s-1)} $ is an operator of  variable $x$,  external coupling weight $\omega<0$, and  internal   coupling weight coefficient $c_l>0$ with $l =0,1,2,\cdots,s-1$.
 Consensus is called to be reached for high-order MAS if $x_{i }^{(\ell)} \rightarrow x_{j}^{(\ell)}$, $\ell=1,2,\dots,s$, $\forall i \neq j$, $i,j=1,2,\cdots,n$.
 The $\ell$-order states of all agents are denoted by a  vector $X^{(\ell)}:=\left[ x_1^{(\ell)}, x_2^{(\ell)},\dots,x_n^{(\ell)}\right]\t, \ell =1,2,\cdots,s $. The full-state of the $i$-th agent is denoted by a  vector 
\begin{equation}
	X_i:=\left[ x_i^{(1)}, x_i^{(2)},\dots,x_i^{(s)}\right]\t , i=1,2,\cdots,n. \label{eq: xi}
\end{equation}

 \vspace{-1em} \hspace{0.2 in}Substituting (\ref{ui}) into (\ref{dynamic}), the dynamics of the whole MAS~$\mathcal G$ can be rewritten in a compact form as 
\begin{equation}
	X(k) = W X(k-1), \label{xk}
\end{equation}
with $X:=\left[  (X^{(1)})\t ,(X^{(2)})\t ,\dots,(X^{(s)})\t  \right]\t $ and \vspace{-1em}
\begin{equation}
	\setlength{\arraycolsep}{2.5pt} 
	W=\left[\begin{array}{cccccc}
		I_n & \epsilon I_n & \mathbf{0}_n & \ldots & \mathbf{0}_n & \mathbf{0}_n \\
		\mathbf{0}_n & I_n & \epsilon I_n & \ldots & \mathbf{0}_n & \mathbf{0}_n \\
		\vdots & \vdots & \vdots & \ddots & \vdots & \vdots \\
		\mathbf{0}_n & \mathbf{0}_n & \mathbf{0}_n & \ldots & I_n & \epsilon I_n \\
		\omega c_{0}\mathcal{L} & \omega c_1 \mathcal{L} & \omega c_2\mathcal{L} & \ldots & \omega c_{s-2}\mathcal{L} & I_n+\omega c_{s-1}\mathcal{L}
	\end{array}\right] \nonumber
\end{equation}
being the Perron matrix \citep{olfati2004consensus}. 

 \vspace{-1em} \hspace{0.2 in}To facilitate further analysis, the following assumptions are necessary.
 \vspace{-1em} 
\begin{assumption}
	MAS $\mathcal{G}=\left( \mathcal{V}, \mathcal{E}, \mathcal{A}\right)$ has a directed spanning tree. \label{assumption}
\end{assumption}\vspace{-1em}
\begin{assumption}
	All the eigenvalues  of Perron matrix $W$ of (\ref{xk}) lie inside the unit circle, i.e., $ \left\|  \lambda_{(W,i)} \right\| \leq 1$, $\lambda_{(W,i)}$ denotes the $i$-th eigenvalue of $W$. \label{assumption2}
\end{assumption}
 \vspace{-1em} \hspace{0.2 in}Assumptions \ref{assumption} and  \ref{assumption2} ensure the consensus accessibility of   the discrete-time high-order linear MAS (\ref{xk}) \citep{wieland2008consensus}. Moreover, we give some definitions to facilitate further derivation.
 \vspace{-1em}
 \begin{definition}
 	\textit{{(Consensus vector)}}  For a discrete-time high-order linear MAS (\ref{xk})  fulfilling Assumptions~\ref{assumption} and \ref{assumption2}, the consensus vector \citep{ren2007high} is denoted by $\wp_{c}(k)$   that satisfies  
 		\begin{equation}
 			\wp_{c}(k) :=\Upsilon(k,\mathcal{L})X(0)  \in \mathbb{R}^{s},
 		\end{equation}
 	 where
 		\begin{small}
 			\begin{equation}
 				\Upsilon(k,\mathcal{L})=
 				\left[\begin{array}{cccc}
 					\mathbf{1}_{n}p\t & k \epsilon  \mathbf{1}_{n}p\t  & \dots & \frac{1}{(s-1)!} (k \epsilon)^{(s-1)}   \mathbf{1}_{n}p\t \\
 					\mathbf{0}_{n} &   \mathbf{1}_{n}p\t &  \dots & \frac{1}{(s-2)!} (k \epsilon)^{(s-2)}   \mathbf{1}_{n}p\t\\
 					\mathbf{0}_{n} & \mathbf{0}_{n}  & \ddots & k \epsilon   \mathbf{1}_{n}p\t \\
 					\mathbf{0}_{n} &  \mathbf{0}_{n} & \dots &    \mathbf{1}_{n}p\t\\
 				\end{array}\right], \label{Upsilon}
 			\end{equation} 
 		\end{small} $p \in \mathbb{R}^{n}$ is a nonnegative vector such that $p\t \mathcal{L}=\mathbf{0}_{n}$ and $p\t \mathbf{1}_{n}=1$,  $\wp_{c}^{(\ell)}(k)=(e_{s}^{\ell })\t\wp_{c}(k)$ denotes the   $\ell$-order consensus value, $k \in \mathbb{N}$ denotes the time instant.
 	\label{consensus-vector}
 \end{definition}
\begin{definition}
	\textit{(Individual disagreement vector)}  For a discrete-time high-order linear MAS (\ref{xk})  fulfilling Assumptions~\ref{assumption} and \ref{assumption2},  the individual disagreement vector is denoted by $\wp_{i}(k)$   that satisfies  
		\begin{equation}
			\wp(k) := X(k)-\wp_{c}(k) \otimes \mathbf{1}_{n},  \label{eq:dis}
		\end{equation}
		where    $\wp(k): = \left[\wp_{1}(k),\wp_{2}(k),\dots,  \wp_{n}(k)\right]\t \in \mathbb{R}^{sn}$ denotes disagreement vector, $$\wp_{i}(k): = \left[\wp_{i}^{(1)}(k),\wp_{1}^{(2)}(k),\dots,  \wp_{i}^{(s)}(k)\right]\t \in \mathbb{R}^{s},$$ $\wp_{i}^{(\ell)}(k)$ refers to  $\ell$-order disagreement value  for $i$-th agent.
	\label{de-disagree}
\end{definition}
 \vspace{-1em}
\begin{definition}
	\textit(Consensus-window-launch-time)  For a discrete-time high order linear MAS (\ref{xk})  fulfilling Assumptions~\ref{assumption} and \ref{assumption2},  given  an error thresohold $\sigma \in \mathbb{R}^+$, the {\it consensus-window-launch-time} is a  time $T_c$ that satisfies 
	\begin{equation}
		T_c=\sup\left\lbrace T\left|  \sum_{j=1}^{s}\sum_{i=1}^{n} \left\|\wp_i^{(j)}\left( \frac{T}{\epsilon}\right) \right\|\leq  \sigma \right. \right\rbrace \label{Tc}
	\end{equation}
	with $\wp_{i}^{(j)}(k)$ given in Definition \ref{de-disagree}, $ T_c \in \mathbb{R}^+$.\label{Consensus-window-launch-time}
\end{definition}
 \vspace{-1em}
\begin{definition}
	\textit{(Principal Minor)}  \citep{RA2012matrix} For a square matrix  $\ast\in \mathbb{R}^{n \times n}$, a $m \times m$ $ (n>m)$ submatrix of matrix $\ast$ is a matrix obtained by deleting arbitrary $n-m$ rows and the corresponding columns. The determinant of the $m \times m $ principal submatrix is named the principal minor of $\ast$.
\end{definition}
 \vspace{-1em}
\begin{definition}
	\textit{(Sum of Principal Minor) } \citep{RA2012matrix} For a square matrix  $\ast\in \mathbb{R}^{n \times n}$. The sum of its principal minor of size $m$ (there are $C_n^{m}$ of them)  is denoted by $E_{m}(\ast)$, i.e., for $m=1$, then  $E_1(\ast)=a_{11}+\cdots+a_{n n}$ and $C_n^1=n$; for $m=n$, then  $E_n(\ast)=\operatorname{det} (\ast)$ and $C_n^n=1$. 
\end{definition}
 \vspace{-1em}
\begin{definition}  \textit{(Minimal Polynomial Pair)} \citep{yuan2013} The minimal polynomial pair of  matrix $W$ is the  monic polynomial of smallest degree $D_i$ that satisfies  $(e_{sn}^{i})\t  q_i(W)=0$ with $q_i(t) := t^{D_i+1}+\sum_{j=0}^{D_i} \alpha_{i,j} t^{j},i=1,2,\dots,sn$. \label{Minimal Polynomial Pair}
\end{definition}
 \vspace{-1em}
 
\begin{remark}
	For each agent $i$ ($i=1,2,\cdots,n$) in an MAS, its own individual state $x_i$ is generally available. Therefore, without loss of generality, consider the case that the measurable output  is $x_i^{(1)}(k)$. The virtual of the present predictor lies in retrieving the global initial state information and Laplacian matrix of a whole MAS indirectly to solve both Problems~1 and 2 merely by using an individual state series $x_i^{(1)}(k)$ with a minimal length $2\overline{D}_i+1$, i.e., an individual memory vector 
	 \vspace{-2em} 
	\begin{small}
		\begin{equation}
			_{0}^{2\overline{D}_i}x_i^{(1)}:=\left[ x_i^{(1)}(0), x_i^{(1)}(1),\ldots, x_i^{(1)}(2\overline{D}_i)\right] \in\mathbb R^{2 \overline{D}_i+1}\label{eq: yi} 
		\end{equation}
	\end{small} with $\overline{D}_i \in \mathbb{Z}^+$ to be determined later. 
\end{remark}

 \vspace{-1em} \hspace{0.2 in}Now, we are ready to introduce the two main problems  addressed in this study as below.
 \vspace{-1em}
\begin{problem}
	\textit{(Minimal-time deadbeat consensus prediction)}  Given a discrete-time high-order linear MAS (\ref{xk})  fulfilling Assumptions~\ref{assumption} and \ref{assumption2},  for some  $i=1,2,\cdots,n$, find  a consensus vector $\widetilde{\wp}_c(k) \in \mathbb{R}^{s}$  that satisfies
	\begin{equation}
		{ \widetilde{\wp}_c\left( k\right)      -\wp_c(k) =\textbf{0}_{s} }
	\end{equation} by the memory vector $_{0}^{2\overline{D}_i}x_i^{(1)}$ given in (\ref{eq: yi}) and $\wp_c(k)$ defined in Definition \ref{consensus-vector}.
	\label{problem1}
\end{problem}
 \vspace{-1em}
\begin{problem}
	\textit{(Instant individual disagreement degree prediction)} Given the discrete-time high-order linear MAS (\ref{xk}), for some  $i=1,2,\cdots,n$, find a individual disagreement vector   $\widetilde\wp_i(k) \in \mathbb{R}^{s}$, such that 
	\begin{equation}
		{ \widetilde\wp_i\left( k\right) - \wp_i(k)= \textbf{0}_{s}} \label{eq:problem2}
	\end{equation} by memory vector $_{0}^{2\overline{D}_i}x_i^{(1)}$ given in (\ref{eq: yi}), $\wp_i(k)$  defined in Definition \ref{de-disagree}.		\label{problem2}
\end{problem} 

 \vspace{-0.5em}
\section{Main technical result }
 \vspace{-0.5em} \hspace{0.2 in}In this section, Problems~1 and 2 will be addressed for the discrete-time high-order linear MAS~(\ref{xk}). To this end, we provide the following Lemmas in advance.
 \vspace{-1em}
\begin{lemma} \citep{sundaram2007finite} The minimal polynomial matrix pair  $q_i(t)$  divides the minimal polynomial  of $W$ for all $i=1,2,\cdots,sn$. \label{divide}
\end{lemma}
 \vspace{-1em}
\begin{lemma}
	$W$ in (\ref{xk}) has exactly $s$ eigenvalues at 1 with geometric	multiplicity of 1. \label{eig1}
\end{lemma}
 \vspace{-1em}
\bproof
Rewriting (\ref{dynamic}) in a compact form, one has that
\begin{equation}	
	X(k) =	\left(   \epsilon C \otimes I_n +  \omega R \otimes \mathcal{L} +I_{sn} \right)  X(k-1)\label{xk2}, 
\end{equation}
with 
 \vspace{-2em}
\begin{small} \begin{align*}
		C&=\left(\begin{array}{cccc}
			0 & 1 & \cdots & 0 \\
			\vdots & \ddots & \ddots & \vdots \\
			0 & 0 & \ddots & 1 \\
			0 & 0 & \cdots & 0
		\end{array}\right)_{s\times s},
		R&=\left(\begin{array}{cccc}
			0 & 0 & \cdots & 0 \\
			\vdots & \ddots & \cdots & 0 \\
			c_{0} & c_1 & \cdots & c_{s-1}
		\end{array}\right)_{s \times s}.
\end{align*}\end{small}

 \vspace{-1em} \hspace{0.2 in}Let $ \Lambda:=\operatorname{diag}\left(\lambda_1, \ldots, \lambda_n\right)$ be a matrix with diagonal elements being the eigenvalues associated with Laplacian $\mathcal{L}$, i.e., there exists a nonsingular matrix $P$ such that $P^{-1} \mathcal{L}P=\Lambda $. Then, multiply both sides of (\ref{xk2}) by $ I_{s}\otimes P^{-1}$, one has 
\begin{align*}	
	\left( I_{s}\otimes P^{-1}\right) X(k) &=(B+I_{sn}) 	\left( I_{s}\otimes P^{-1} \right)X(k-1)
\end{align*}
with $B=\left( \epsilon C \otimes I_n +\omega  R \otimes \Lambda    \right)  $.
Let $\eta(k)=\left(I_{s} \otimes P^{-1} \right) X(k)$, one has
$$
{\eta}_i(k) = \left( B_i+I_{s} \right) \eta_i(k-1)
$$
with $\eta_i(k)=\left[e_{s n}^i , e_{s n}^{n+i}, \cdots ,e_{s n}^{(s-1)n+i } \right]\t \eta(k)$, $B_i=\left(  \epsilon	C +\omega \lambda_i R\right)$ and $i = 1,2,\dots,n $.

 \vspace{-1em} \hspace{0.2 in}According to \citep{RA2012matrix}, the characteristic polynomial of $B_i$ is  derived as follows\vspace{-2em}
 \begin{small}
\begin{align*}
	p_{B_i}(\lambda)
	=&\lambda^{s}-\omega c_{s-1} \lambda_i \lambda^{s-1}- \omega c_{s-2}\epsilon\lambda_i \lambda^{s-2}- \omega c_{s-3}\epsilon^2\lambda_i \lambda^{s-3} \\
	&-\cdots-\omega c_1  \epsilon^{s-2} \lambda_i \lambda-\omega c_{0} \epsilon^{s-1}\lambda_i.
\end{align*}
\end{small}

 \vspace{-1em} \hspace{0.2 in}Since $\mathcal{L}$ has a simple zero eigenvalue, i.e., there exists one and only one $\lambda_i=0$. 
Consequently, $	p_{B_i}(\lambda)$  has  $s$ zero eigenvalues, which implies that  $B+I_{sn}$ only has $s$ eigenvalues at 1. Moreover, 
$$
W = 	\left( I_{s}\otimes P^{-1} \right)^{-1} \left( B+I_{sn} \right)\left( I_{s}\otimes P^{-1} \right),
$$
which implies that $W$ has only $s$ eigenvalues at 1 as well.

 \vspace{-1em} \hspace{0.2 in}Let $\mathcal{Q} := [\mathcal{Q}_1\t  , \mathcal{Q}_2\t , \dots \mathcal{Q}_s\t ]\t  \in \mathbb{R}^{sn}$ denotes the right eigenvector of $W$ associated to the eigenvalue 1, then one has
$$
W\mathcal{Q}=\mathcal{Q},
$$
which implies $\mathcal{Q}_i=\mathbf{0}_n,i = 2,3\dots s$ whereas $\mathcal{L}\mathcal{Q}_1=\mathbf{0}_n$. Thereby, $\mathcal{Q}_1$ is an eigenvector of $\mathcal{L}$ associated to its eigenvalue at 0. Since $\mathcal{L}$ has only one linearly independent eigenvector $\mathcal{Q}_1$ associated to the  eigenvalue at 0, which in turn implies that $W$ has only one linearly independent eigenvector $\mathcal{Q}=\left[\mathcal{Q}_1\t ,\mathbf{0}_n\t ,\dots,\mathbf{0}_n\t   \right]\t $. Therefore, the  eigenvalue of $W$ at 1 has algebraic multiplicity of $s$ but geometric multiplicity of 1. The proof is
thus completed.	\eproof
 \vspace{-1em}
\begin{lemma}
	The minimal polynomial pair  of matrix $W$  can be decomposed into $q_i(t)=(t-1)^{\ell(i)}p_i$, where $\ell(i)=s-\lfloor \frac{i}{n}\rfloor, i=1,2\cdots, sn$; $p_i$ denotes a polynomial without the factor $(t-1)$.
	\label{seigenvalue}
\end{lemma} \vspace{-1em}
\bproof
Let $\lambda$ be an eigenvalue of $W$ that the algebraic multiplicity of $s$ and geometric	multiplicity of 1. Define a Jordan block of order $s$ as follows
$$
\boldsymbol{J}(\lambda):=\left(\begin{array}{ccccc}
	\lambda & 1 & & & \\
	& \lambda & 1 & & \\
	& & \ddots & \ddots & \\
	& & & \ddots & 1 \\
	& & & & \lambda
\end{array}\right)_{s \times s}.
$$

 \vspace{-1em} \hspace{0.2 in}One has that
  \vspace{-2em}
\begin{small}
	$$
	\begin{aligned}
		\setlength{\arraycolsep}{2pt} 
		\boldsymbol{J}^{k}(\lambda)=
		\left\lbrace 
		\begin{split}
			&\left(\begin{array}{crclc}
				\lambda^{k} & C_k^1 \lambda^{k-1} & \cdots & 1 & 0 \\
				& \lambda^{k} \qquad\quad & C_k^1 \lambda^{k-1} & \text{  }\ddots & 0 \\
				& \ddots & \qquad\ddots &  & 1 \\
				& & \ddots & & \\
				& & & & C_k^1 \lambda^{k-1} \\
				& & & & \lambda^{k}
			\end{array}\right),k<s
			\\
			&\left(\begin{array}{cclc}
				\lambda^{k} & C_k^1 \lambda^{k-1} & \quad \cdots & C_k^{s-1} \lambda^{k-s+1} \\
				& \lambda^{k} & C_k^1 \lambda^{k-1} & \vdots \\
				& & \ddots & \\
				& & & C_k^1 \lambda^{k-1} \\
				& & & \lambda^{k}
			\end{array}\right) ,k\geq s
		\end{split}\right.
	\end{aligned}.
    $$
\end{small}

 \vspace{-1em} \hspace{0.2 in}Case 1: if $\ell(i)=1$, one has that
\begin{equation}
	q_{i,1}(\lambda)={q_i(\lambda)}=
	\lambda^{D_i+1}+\alpha_{i,D_i} \lambda^{D_i}+\cdots+\alpha_{i,1} \lambda+\alpha_{i,0}=0. \label{qi1}
\end{equation}

 \vspace{-1em} \hspace{0.2 in}Case 2:  if $\ell(i)=m$ with $ m \in \mathbb{Z}^+, 2 \leq m \leq s$. By mathematical induction,
one has that
\begin{equation}
	q_{i,m}(t)=\frac{q_i(t)}{(t-\lambda)^{m-1}}=\sum_{j=0}^{{D_i}-m+2}v_{j,m}t^j\label{qim1}
\end{equation}
 with 			$v_{j,m}=\sum_{h=j}^{{D_i}-m+2}C_{m-2+h-j}^{m-2}\alpha_{i,h+m-1}\lambda^{h-j}$.

 \vspace{-1em} \hspace{0.2 in}Bearing in mind Pascal's Triangle $\sum_{j=0}^xC_{m-1+j}^{m-1}=C_{m+x}^{m},x\in \mathbb{Z}^+$, substituting $t=\lambda$ into (\ref{qim1}) yields
\begin{equation}
	q_{i,m}(\lambda)=\sum_{j=m-1}^{D_i+1}\alpha_{i,j}C_j^{m-1}\lambda^{j-m+1}=0, \label{qim}
\end{equation}
and $q_{i,x}(\lambda)=0,x\in \mathbb{Z}^+,x<m$.

 \vspace{-1em} \hspace{0.2 in}Combining Cases~1 and 2, one has that $q_{i,x}(\lambda)=0,x\in \mathbb{Z}^+,x \leq \ell(i)$. According to Lemma~\ref{eig1}, let the Jordan decomposition of $W$ be indicated by
$$
W=\mathcal{T}\left[\begin{array}{cc}
	\boldsymbol{J}(1) & \textbf{0} \\
	\textbf{0} & Z
\end{array}\right] \mathcal{T}^{-1},
$$
where $Z$ denotes the set of Jordan block whose eigenvalue does not equal 1, $\mathcal{T}$ is a nonsingular matrix with columns composed of the (generalized) eigenvectors of $W$.

 \vspace{-1em} \hspace{0.2 in}From Definition \ref{Minimal Polynomial Pair}, one has 
\begin{equation}
	(e_{sn}^{i})\t  \mathcal{T}\left[\begin{array}{cc}
		q_i(\boldsymbol{J}(1)) & \textbf{0} \\
		\textbf{0} & q_i(Z )
	\end{array}\right] \mathcal{T}^{-1}=0.
\end{equation}

 \vspace{-1em} \hspace{0.2 in}Let $\mathbf{d} \in \mathbb{R}^{sn\times s}$ denotes the  $1$-th to $s$-th columns of $ \mathcal{T}$,  direct calculation gives 
\begin{equation}
	[_{1}^{s}d_{i} \quad _{s+1}^{sn}\mathcal{T}_{i}]\left[\begin{array}{cc}
		q_i(\boldsymbol{J}(1)) & \textbf{0} \\
		\textbf{0} & q_i(Z)
	\end{array}\right]=0 ,\label{dT}
\end{equation}
where  $_{1}^{s}d_{i}$ is $i$-th row of $\mathbf{d}$ and $[_{1}^{s}d_{i} \quad _{s+1}^{sn}\mathcal{T}_{i}]$ is the $i$-th row of $ \mathcal{T}$. According to Lemma~\ref{eig1}, $W$ has only one linearly independent eigenvector $\mathcal{Q}=\left[\mathcal{Q}_1\t ,\mathbf{0}_n\t ,\dots,\mathbf{0}_n\t   \right]\t$ associated to eigenvalue 1. Without loss of generality, let $\mathcal{Q}=\left[\mathbf{1}_n\t ,\mathbf{0}_n\t ,\dots, \mathbf{0}_n\t \right]\t $ and $ \mathcal{Q}_{1,r}=\left[\mathbf{0}_n\t ,\mathbf{0}_n\t ,\dots,\frac{1}{\epsilon^{r-1}}\mathbf{1}_{r\text{-}th}\t , \cdots \mathbf{0}_n\t \right]\t  , r = 2,3\dots,s$, be a right eigenvector and generalized right eigenvectors of $W$ associated to the  eigenvalue at 1, respectively. Then,
$$
\mathbf{d}=\left[\begin{array}{ccccc}
	\mathbf{1}_n & \textbf{0}_{n} & \textbf{0}_{n} & \cdots & \textbf{0}_{n} \\
	\textbf{0}_{n} & \frac{1}{\epsilon}\mathbf{1}_n  & \textbf{0}_{n} & \cdots & \textbf{0}_{n} \\
	\textbf{0}_{n} & \textbf{0}_{n} & \frac{1}{\epsilon^2}\mathbf{1}_n  & \cdots & \vdots \\
	\vdots & \vdots & \vdots & \ddots & \vdots \\
	\textbf{0}_{n} & \textbf{0}_{n} & \textbf{0}_{n} & \cdots & \frac{1}{\epsilon^{s-1}} \mathbf{1}_n 
\end{array}\right].
$$

 \vspace{-1em} \hspace{0.2 in}Bearing  $	\boldsymbol{J}^{k}(\lambda)$ and  (\ref{qi1}), (\ref{qim}) in mind, one has
$$
q_i(\boldsymbol{J}(1))=\left[\begin{array}{cccc}
	q_{i, 1}(1) & q_{i, 2}(1) & \ldots & q_{i, s}(1) \\
	0 & q_{i, 1}(1) & \cdots & q_{i , s-1}(1) \\
	\vdots & \vdots & \cdots & \vdots \\
	0 & 0 & \cdots & q_{i, 1}(1)
\end{array}\right].
$$

 \vspace{-1em} \hspace{0.2 in}From (\ref{dT}), one has $_{1}^{s}d_{i}q_i(J(1))=0$, all elements of $\left( \lfloor \frac{i}{n}\rfloor+1\right) $-row of $q_i(J(1))$ are zeros for $i=1,2,\dots,sn$, i.e., $q_{i,m}(1)=0,m= 1,2,\cdots,\ell(i)$, and the minimal polynomial pair $q_i(t)$ of $W$ has at least $\ell(i)$ eigenvalues at $1$. Since the degree of the  polynomial pair is minimized,  $q_i(t)$ has and only has $\ell(i)$ eigenvalues at 1.  The  proof is thus completed.	\eproof  \vspace{-1em}
\begin{remark} 
	The number of eigenvalue 1 contained in the minimal polynomial pair  of matrix $W$ could be determined by a Hankel matrix construction, which will be designed  later.
\end{remark}\vspace{-1em}
\begin{remark}
	Note that $\mathcal{Z}(C_k^{r})=\frac{z}{(z-1)^{r+1}},r\in \mathbb{N}$, $k$ denotes  the $k$-th time instant. Firstly,  $\mathcal{Z}(C_k^{0})=\frac{z}{z-1}$ for $r=0$. Suppose $\mathcal{Z}(C_k^{m})=\frac{z}{(z-1)^{m+1}}$ holds for  $r=m$, one has $\mathcal{Z}(C_k^{m+1})=\mathcal{Z}\left( \frac{k-m}{m+1}C_k^{m}\right) =\frac{z}{(z-1)^{m+2}}$ for $r=m+1$.  	\label{ckr}
\end{remark} \vspace{-1em}
\begin{thm}
	Consider   a closed-loop high-order MAS  governed by   (\ref{dynamic}), (\ref{ui}).  Under Assumptions \ref{assumption} and  \ref{assumption2},  Problem \ref{problem1}  is solvable. 
\end{thm} \vspace{-1em}

\bproof
The proof can be divided into three steps.

Step~1: \textit{Derive the parameter of the minimal polynomial pair.}

 \vspace{-1em} \hspace{0.2 in}Consider the memory vector (\ref{eq: yi}) and Lemma~\ref{seigenvalue}, one has 
\begin{equation}
	q_i(t)=(t-1)^{s}p_i(t)=\sum_{j=0}^{D_i+1}\alpha_{i,j}t^{j} \label{qit}
\end{equation} with $\alpha_{i}=\left[ \alpha_{i,0},\alpha_{i,1},
\dots,\alpha_{i,D_{i}+1}\right] \in \mathbb{R}^{D_{i}+2}$ being  the coefficients of the minimal polynomial pairs of $W$ to be calculated.  Note that
\begin{equation}
	p_i(t) = \frac{q_i(t)}{(t-1)^{s}} = \sum_{j=0}^{D_i+1-s}\zeta_{i,j}t^{j}, \label{p_i}
\end{equation}
where $\zeta_{i,j} \in \mathbb{R}$ is another null space coefficient parameters. From  Leibniz's Derivation Rule \citep{olver1993applications} and Definition \ref{Minimal Polynomial Pair}, one has 
\begin{equation}
	\zeta_i \Theta_i(k,\overline{D}_i) = 0 \label{prehankel}
\end{equation}
with $\zeta_i = \left[\zeta_{i,0}, \zeta_{i,1} ,\dots, \zeta_{i,\overline{D}_i} \right],\overline{D}_i=D_i+1-s$ and \vspace{-2em}
\begin{small}
	\begin{equation}
		\Theta_i(k,\overline{D}_i) = \left[\begin{split}
			&\sum_{h=0}^{s}(-1)^{s-h}C_{s}^{h}x_{i}^{(1)}(k+h)&\\
			&\sum_{h=0}^{s}(-1)^{s-h}C_{s}^{h}x_{i}^{(1)}(k+h+1)& \\
			&\qquad\qquad\qquad\vdots&\\
			&\sum_{h=0}^{s}(-1)^{s-h}C_{s}^{h}x_{i}^{(1)}(k+h+\overline{D}_i)&
		\end{split}\right].
	\end{equation}
\end{small}

 \vspace{-1em} \hspace{0.2 in}According to (\ref{prehankel}), build up a Hankel matrix   \citep{partington1988introduction} as the following form\vspace{-2em}
\begin{small}
	\begin{equation}
		\begin{split}
			&\Gamma=\left[\begin{array}{cccc}
				\Theta_i(k,\widehat{\overline{D}}_i) & \Theta_i(k+1,\widehat{\overline{D}}_i) & \cdots & \Theta_i(k+\widehat{\overline{D}}_i,\widehat{\overline{D}}_i) \\
			\end{array}\right].\label{Gamma}
		\end{split}
	\end{equation}
\end{small} 
Gradually increase $\widehat{\overline{D}}_i$  until getting the minimal $\widehat{\overline{D}}_i$ such that   the Hankel matrix $\Gamma$ becomes non-fully ranked. Then, $\widehat{\overline{D}}_i$ and  $\Gamma_i^\bot$  become  estimates of $\overline{D}_i$ and $\zeta_i$, respectively. Together with (\ref{qit}), (\ref{p_i}), $\alpha_{i,j}$ can be obtained.

 \vspace{-1em} \hspace{0.2 in}Step~2: \textit{Calculate and decompose-fit the $\mathcal{Z}$-transform of $x_{i}^{(1)}(t)$}.

 \vspace{-1em} \hspace{0.2 in}According to (\ref{qit}) and Definition \ref{Minimal Polynomial Pair}, one has   \vspace{-2em}
 \begin{small}
\begin{equation}
	\begin{split}
		\left( e_{sn}^{i}\right) \t\sum_{j=0}^{D_i+1}\alpha_{i,j}W^{j}X(k)
		&=\sum_{j=0}^{D_i+1}\alpha_{i,j}x_{i}^{(1)}(k+j)=0.
	\end{split} \label{gammak}
\end{equation}
 \end{small}

 \vspace{-1em} \hspace{0.2 in}Note that the $\mathcal{Z}$-transform $x_{i}^{(1)}(z)=\mathcal{Z}\left(x_{i}^{(1)}(t)\right)$, from $\mathcal{Z}$-transform time-shift properties and  (\ref{gammak}), one has 
\begin{equation}
	\begin{split}
		x_{i}^{(1)}(z)& = \frac{\sum_{j=1}^{D_i+1} \alpha_{i,j}z^{j}\left(\sum_{k=0}^{j-1} x_{i}^{(1)}(k) z^{-k}\right)}{(z-1)^{s}p_i(z)} \\
		&=\frac{\sum_{j=1}^{D_i+1} z^{j} \sum_{h=0}^{D_i+1-j} \alpha_{i,h+j} x_{i}^{(1)}(h)}{(z-1)^{s} p_i(z)}.\\
	\end{split}
	\label{Xz}
\end{equation}

 \vspace{-1em} \hspace{0.2 in}Let 
\begin{equation}
	\phi_i(z)=\sum_{j=0}^{D_i} z^{j} \sum_{h=0}^{D_i+1-j} \alpha_{i,h+j} x_{i}^{(1)}(h),\label{Phi1}
\end{equation}
which could be divided into two parts as 
\begin{equation}
	\phi_i(z)= p_i(z)\sum_{j=0}^{s-1}\beta_j (z-1)^{j} + \sum_{j=s}^{D_i} \beta_{j	}(z-1)^{j} .\label{Phi2}
\end{equation}
Here, $\beta_i \in \mathbb{R},i=0,1,\cdots, D_i$, is a parameter.

 \vspace{-1em} \hspace{0.2 in}From  Leibniz's Derivation Rule \citep{olver1993applications}, the $n$-th derivative of (\ref{Phi2}) is given as follows
\begin{equation}
	\phi_i^n(1) = \sum_{k=0}^nA_n^{k}p_i^{n-k}(1)\beta_k,
\end{equation}
for $ n= 0,1,\cdots, s-1$. Direct calculation gives 
\begin{equation}
	\left[\begin{array}{c}
		\beta_{0} \\
		\beta_1 \\
		\vdots \\
		\beta_{s-1}
	\end{array}\right]=
	\mathcal{M}^{-1}
	\left[\begin{array}{c}
		\phi_i^{(0)}(1) \\
		\phi_i^{(1)}(1) \\
		\vdots \\
		\phi_i^{(s-1)}(1)
	\end{array}\right],\label{beta}
\end{equation}
with  \vspace{-2em} 
\begin{small}
	\begin{align*}
		\setlength{\arraycolsep}{2.5pt} 
		\mathcal{M}=\left[\begin{array}{ccccc}
			p_i^{0}(1) & 0 & 0 & \cdots & 0 \\
			p_i^1(1) & p_i^{0}(1) & 0 & \cdots & 0 \\
			p_i^2(1) & A_2^1 p_i^1(1) & A_2^2 p_i^{0}(1) & \cdots & 0 \\
			\vdots & \vdots & \vdots & \ddots & \vdots \\
			p_i^{s-1}(1) & A_{s-1}^1 p_i^{s-2}(1) & A_{s-1}^2 p_i^{s-3}(1) & \cdots &  A_{s-1}^{s-1}p_i^{0}(1)
		\end{array}\right].
	\end{align*}
\end{small}

 \vspace{-1em} \hspace{0.2 in}Since $p_i^{0}(1) \neq 0$,  $\mathcal{M}$ is fully ranked, and $\beta_i$ always exists such that (\ref{Phi2}) holds.

 \vspace{-1em} \hspace{0.2 in}Step~3: \textit{Predict the eventual consensus value of the MAS.}

 \vspace{-1em} \hspace{0.2 in}Under Assumptions \ref{assumption} and  \ref{assumption2},  $x_i^{(s)}$ converges to a constant value.  Denote the consensus item as the following form
\begin{equation}
	 \widetilde{\wp}_c^{(1)}(k)= \kappa_{s-1}k^{s-1}+\kappa_{s-2}k^{s-2}+\dots+\kappa_{0}k^{0}, \label{Ft}
\end{equation}
where $\kappa_i,i =0,1,\dots,s-1,$ is the  coefficient to be determined later. According to the property of  $\mathcal{Z}(C_k^{r})=\frac{z}{(z-1)^{r+1}},r \in \mathbb{N}$, in Remark \ref{ckr}, rewriting  (\ref{Ft})  in a compact form as
\begin{equation}
	\Psi(k) = \sum_{r=0}^{s-1}b_rC_k^{r},\label{Psi}
\end{equation}
where $b_r$ is the coefficient to be determined.

 \vspace{-1em} \hspace{0.2 in}Substituting (\ref{Ft}) into (\ref{Psi}) yields
\begin{equation}
	\left[\begin{array}{c}
		\kappa_{s-1} \\
		\kappa_{s-2} \\
		\vdots \\
		\kappa_{0}
	\end{array}\right]=
	\mathcal{J}^{-1}
	\left[\begin{array}{c}
		\Psi^{(s-1)}(1) \\
		\Psi^{(s-2)}(1) \\
		\vdots \\
		\Psi^{(0)}(1)
	\end{array}\right]\label{c}
\end{equation}
with
\begin{align*}
	\mathcal{J}=\left[\begin{array}{ccccc}
		A_{s-1}^{s-1} & 0 & 0 & \cdots & 0 \\
		A_{s-1}^{s-2} & A_{s-2}^{s-2} & 0 & \cdots & 0 \\
		A_{s-1}^{s-3} & A_{s-2}^{s-3} & A_{s-3}^{s-3} & \cdots & 0 \\
		\vdots & \vdots & \vdots & \ddots & \vdots \\
		A_{s-1}^{0} & A_{s-2}^{0} & A_{s-3}^{0}  & \cdots &  A_{0}^{0}
	\end{array}\right].
\end{align*}

 \vspace{-1em} \hspace{0.2 in}Let  $\wp_c^{(1)}(z)=\mathcal{Z}\left( \widetilde{\wp}_c^{(1)}(k)\right)$, one has
\begin{equation}
	 \widetilde{\wp}_c^{(1)}(z)=\sum_{r=0}^{s-1}b_r\frac{z}{(z-1)^{r+1}}.
	\label{Fz}
\end{equation}

 \vspace{-1em} \hspace{0.2 in}Let  $\widetilde\wp_i^{(1)}(z)$ denote the disagreement item  of $x_{i}^{(1)}(z)$ , from   (\ref{Xz}), (\ref{Fz}), one has  \vspace{-2em}
 \begin{small}
\begin{equation}
	\begin{split}
		H(z)=&	x_{i}^{(1)}(z) -  \widetilde{\wp}_c^{(1)}(z) - \widetilde\wp_i^{(1)}(z) \\
		=&\frac{z\phi_i(z)}{(z-1)^{s} p_i(z)} -\sum_{j=0}^{s-1}b_j\frac{z}{(z-1)^{j+1}} -\widetilde\wp_i^{(1)}(z)\\
		= &z\frac{\sum_{j=0}^{s-1}\left(  \beta_j-b_{s-1-j}\right) (z-1)^{j} }{(z-1)^{s}}\\
		& + z\frac{ \sum_{j=s}^{D_i} \beta_j(z-1)^{j}}{(z-1)^{s} p_i(z)} - \widetilde\wp_i^{(1)}(z).
	\end{split}\label{Hz}
\end{equation}
 \end{small} \vspace{-1em}

 \vspace{-1em} \hspace{0.2 in}Moreover, it follows from $H(z)=0$ that $b_{s-1-j} = \beta_j \text{ with }j =0,1,\cdots, s-1$ and $\beta_j,j= 0,1, \cdots, s-1$, has been obtained in (\ref{beta}). Then,  $\kappa_j$ can be calculated by (\ref{c}), and hence the consensus item~(\ref{Ft}) is  obtained. 

 \vspace{-1em} \hspace{0.2 in}Direct calculation gives
\begin{equation}
	 \widetilde{\wp}_c^{(j)}(k)= \frac{  \widetilde{\wp}_c^{(j-1)}(k+1)- \widetilde{\wp}_c^{(j-1)}(k)}{\epsilon}\label{sj}
\end{equation}
with $j= 2,3,\cdots,s$. Thereby, all orders of  the consensus item for MAS (\ref{xk}) could be obtained as well. The proof is
thus completed.
\eproof \vspace{-1em}

\begin{thm} 
	Consider   a  closed-loop high-order MAS  governed by   (\ref{dynamic}), (\ref{ui}). Under Assumptions~\ref{assumption} and  \ref{assumption2},   Problem \ref{problem2} is solvable.
\end{thm}\vspace{-1em}

\bproof
Let  $\Omega_{i,1}(z)= p_i(z)\sum_{j=0}^{s-1} \beta_j(z-1)^{j} $,  $\Omega_{i,2}(z)= \sum_{j=s}^{D_i} \beta_{j	}(z-1)^{j}$. From (\ref{Phi2}), $ \Omega_{i,2}(z) = \phi_i(z) - \Omega_{i,1}(z) $, then
\begin{equation}
	\beta_j=\frac{\Omega_{i,2}^{(j)}(1)}{j!},\label{beta2}
\end{equation}
for $j =s,s+1,\cdots, D_i$.

 \vspace{-1em} \hspace{0.2 in}According to  (\ref{Hz}), the  instant individual disagreement value  of $i$, i.e.,   $\widetilde\wp_i^{(1)}(z)$  is determined as follows
\begin{equation}
	\begin{split}
		\widetilde\wp_i^{(1)}(z)&=z\frac{\sum_{j=s}^{D_i} \beta_j(z-1)^{j-s}}{p_i(z)}.
	\end{split}\label{deltaz}
\end{equation}

 \vspace{-1em} \hspace{0.2 in}According to Lemma~\ref{divide} and (\ref{p_i}), $p_i(z)$ can denoted by
\begin{equation}
	p_i(z)=(z-\lambda_1)(z-\lambda_2)\dots(z-\lambda_{\overline{D}_i}),\label{pz}
\end{equation}
where $\lambda_i,i=1,2,\cdots, \overline{D}_i$, is the eigenvalue of matrix $W$.
Therefore, consider the partial fraction expansion of ${\widetilde\wp_i^{(1)}(z)}/{z}$ as below \vspace{-2em}
 \begin{small}
$$
	\begin{aligned}
		\frac{\widetilde\wp_i^{(1)}(z)}{z}=\frac{K_{0}}{z} + \sum_{h \in \mathcal{F}_1 \cup \mathcal{F}_2,\lambda_{h} \neq 0}\frac{K_{h}}{z-\lambda_{h}}+\sum_{h \in \mathcal{F}_3}\sum_{h_i=1}^{\overline{n}_{h}}\frac{K_{h,h_i}}{\left(z-\lambda_{h}\right)^{h_i}},
	\end{aligned}
$$
\end{small}
where   $\mathcal{F}_1,\mathcal{F}_2,\mathcal{F}_3$ indicate single real roots, conjugate single pole and repeated roots sets, respectively, and $\overline{n}_{h}$ is the multiplicity for ${h} \in \mathcal{F}_3$. 

 \vspace{-1em} \hspace{0.2 in}The  partial fraction coefficient for ${h} \in \mathcal{F}_1 \cup \mathcal{F}_2 $ is calculated by
\begin{equation}
	K_{h}=\left.\left(z-\lambda_{h}\right) \frac{\widetilde\wp_i^{(1)}(z)}{z}\right|_{z=\lambda_{h}},\label{Ki}
\end{equation}
for $h \in \mathcal{F}_3$ and 
$$ 
K_{h, h_i}=\left.\frac{1}{(\overline{n}_{h}-h_i) !} \frac{\mathrm{d}^{\overline{n}_{h}-h_i}}{\mathrm{~d} z^{\overline{n}_{h}-h_i}}\left[\left(z-\lambda_{h}\right)^{\overline{n}_{h}} 	\frac{\widetilde\wp_i^{(1)}(z)}{z}\right]\right|_{z=\lambda_{h}}.
$$

 \vspace{-1em} \hspace{0.2 in}Let $\mathcal{F}_2^+$ be a set whose  imaginary part is positive in $\mathcal{F}_2$. For $h \in \mathcal{F}_2^+$,  $\lambda_{h}=\mathrm{m} e^{j\mathfrak{n}}$, $K_{h}=\left| K_{h}\right|e^{j\theta} $, where $j$ denotes the imaginary unit.

 \vspace{-1em} \hspace{0.2 in}Taking the inverse $\mathcal{Z}$-transform as
\begin{equation}
	\begin{split}
		\widetilde\wp_i^{(1)}(k)=&\sum_{{h} \in \mathcal{F}_1,\lambda_{h} \neq 0} 		K_{h}\left(\lambda_{h}\right)^{k} \nu(k) +K_{0}\varepsilon(k)\\
		&+\sum_{{h} \in \mathcal{F}_2} 2\left|K_{h}\right| \mathrm{m}^{k} \cos (\mathfrak{n} k+\theta) \nu(k)\\
		&+\sum_{{h} \in \mathcal{F}_3}\sum_{h_i=1}^{\overline{n}_{h}}\frac{K_{h,h_i}}{(h_i-1) !} \frac{k!}{(k-h_i+1)!} \lambda_{h}^{k-h_i+1} \nu(k),\\
		\label{wpd}
	\end{split}
\end{equation} 	where $\nu(k), \varepsilon(k)$ are step  and impulse functions, respectively.

 \vspace{-1em} \hspace{0.2 in}Analogous to (\ref{sj}), all orders of instant individual   disagreement  $\widetilde\wp_i$, i.e., $\widetilde\wp_{i}^{(1)}, \dots,\widetilde\wp_{i}^{(s)},$ for MAS (\ref{xk}) can be obtained. The proof is
thus completed.
\eproof

 \vspace{-1em} \hspace{0.2 in}Algorithm 1 is presented  below to calculate the eventual consensus value of the MASs, and afterwards the instant individual disagreement value.  
\begin{algorithm}
	\renewcommand{\algorithmicrequire}{\textbf{Input:}}
	\renewcommand{\algorithmicensure}{\textbf{Output:}}
	\caption{MDCP-based instant individual disagreement degree prediction  algorithm for  high-order linear MASs}
	\label{alg1}
	\begin{algorithmic}[1]
		\Require Memory vector $_{0}^{2\overline{D}_i}x_i^{(1)},i=1,2,\cdots,n$, sampling time $\epsilon$.
		\Ensure  The  consensus item $ \widetilde{\wp}_c$, the  instant individual  disagreement item $\widetilde\wp_i$.
		\State Set the order of MAS  $s$, the maximal number of iterations $I_{\max}$. Initialize $\widehat{\overline
			{D}}_i=0$.
		\For{$j=1:I_{\max}$}
		\State Compute the Hankel matrix $\Gamma$ in  (\ref{Gamma}). 
		\If{rank($\Gamma$) $<$ size($\Gamma$)}
		\State Break.
		\EndIf
		\State $\widehat{\overline{D}}_i = \widehat{\overline{D}}_i + 1$
		\EndFor
		\State   Calculate the coefficient $\zeta_i=\Gamma^\bot$ in  (\ref{p_i}), $\overline{D}_i=\widehat{\overline{D}}_i$. Calculate the coefficients $\alpha_{i,j}$ according to (\ref{qit}) and (\ref{p_i}).
		\State Calculate $x_{i}^{(1)}(z)$ and $\phi_i(z)$  according to  (\ref{Xz}), (\ref{Phi2}).
		\State Calculate  $\beta_j$  in (\ref{Phi2}) using  (\ref{beta}), (\ref{beta2}). 
		\State   $b_{s-1-j}=\beta_j,j= 0,1, \cdots, s-1$.
		\State  Calculate $\kappa_i$ using  (\ref{c}). 
		\State Calculate the consensus item $ \widetilde{\wp}_c(k)$ using (\ref{Ft}), (\ref{sj}).
		\State Obtain the instant individual disagreement value $\widetilde\wp_i$ using  (\ref{sj}), (\ref{wpd}).
	\end{algorithmic}  
\end{algorithm}
\vspace{-1em}
\begin{corollary}
	Denote the  state vector $X_i(k)$ in (\ref{eq: xi})   by  $X_i(k) =  \widetilde{\wp}_c(k)+\widetilde\wp_i(k)$.
	Let 
	$$
	W =S J S^{-1},
	$$
	where $J \in \mathbb{R}^{sn \times sn}$ is a diagonal matrix, $S \in \mathbb{R}^{sn \times sn} $ is a nonsingular matrix whose columns are (generalized) eigenvectors of $W$. Let  $\mathcal{K}:=\left[\mathcal{K}_1, \mathcal{K}_2, \ldots, \mathcal{K}_{s}\right]\t $ with $\mathcal{K}_r=$ $S^{-1}_r X(0), r= 1,2,\cdots,s$, $S^{-1}_r$ denoting the $r$-th row of $S^{-1}$. Then, one has  \vspace{-1em}
	\begin{equation}
		\lim _{k \rightarrow \infty} X_i(k)-\sum_{r=1}^{s} e_{s}^{r} \sum_{h=0}^{s-r} C_k^{h} \mathcal{K}_{r+h}=0. \label{Huzhe}
	\end{equation} 
\end{corollary}\label{col: final}  \vspace{-1em}

\bproof
Since  $\lim_{k \rightarrow \infty} \widetilde\wp_i(k)=0$, one has $\lim_{k \rightarrow \infty} X_i(k)= \widetilde{\wp}_{c(k) }$. According to  (\ref{Psi}) and \citep{hu2019decentralized}, it can be derived that $ \widetilde{\wp}_c(k)= \sum_{r=1}^{s} e_{s}^{r} \sum_{h=0}^{s-r} C_k^{h} \mathcal{K}_{r+h}$. 
\eproof

 \vspace{-1em} \hspace{0.2 in}Note that Corollary~1 is a special case corresponding to  $k \rightarrow +\infty$ of Theorem~2.  \vspace{-1em}
\begin{remark}
	The purpose of most consensus prediction methods for a first-order MAS is to recover  the consensus value (e.g., average consensus $\frac{1}{n}\sum_{i}^{n}x_{i}^{(1)}(0) \in \mathbb{R}$).   By constrast, in the present study, we need to recover the eventual consensus vector (i.e., recover $\Upsilon(k,\mathcal{L})X(0)$ given in (\ref{Upsilon})), which is more general yet challenging due to the lack of niche analytical tools. More precisely, the challenge of proposed MDCP lies in the establishment of the quantitative relationship between  the eigenspectrum  of Perron matrix and  the minimal polynomial pair, and decomposition-fit of $\mathcal{Z}$-transform of state. The conventional kind of first-order discrete-time consensus prediction algorithms is a special case of the present MDCP with $s=1$.
\end{remark}
 \vspace{-1em}
 \begin{remark}
 	In this study, the conventional control protocol for high order linear MASs \citep{ren2007high} is employed to reach consensus  in relative damping manner. Since the Perron matrix has a simple eigenvalue at 1 in the absolute damping manner, the present method can be  extended to such a  damping scenario with the assistance of Lemma 1 in  \citep{sundaram2007finite}.
 \end{remark}
\section{Numerical simulation}
\begin{figure}[htbp]
	\centering
	\includegraphics[width=3.6cm]{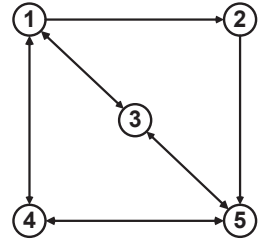}
	\caption{Communication topology graph $\mathcal{G}$.}\label{topology}
\end{figure}
\begin{figure*}[htbp]
	\centering
	\subfigure[Temporal evolution of  $x_{i}^{(4)}$ for routine protocol (a1) and MDCP (a2).]{
		\begin{minipage}[b]{0.46\linewidth}
			\centering
			\noindent\includegraphics[width=0.9\textwidth]{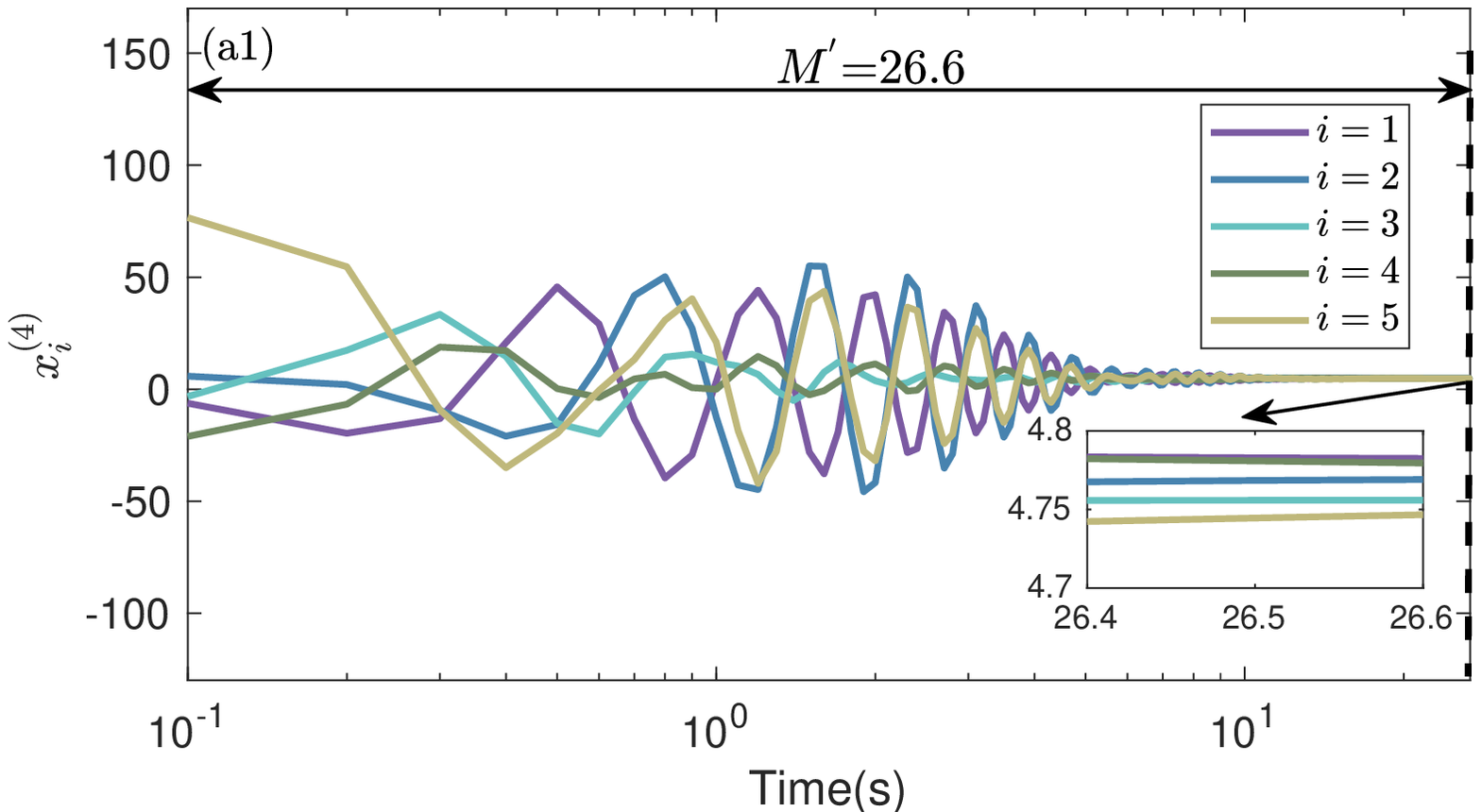}\
			\noindent\includegraphics[width=0.9\textwidth]{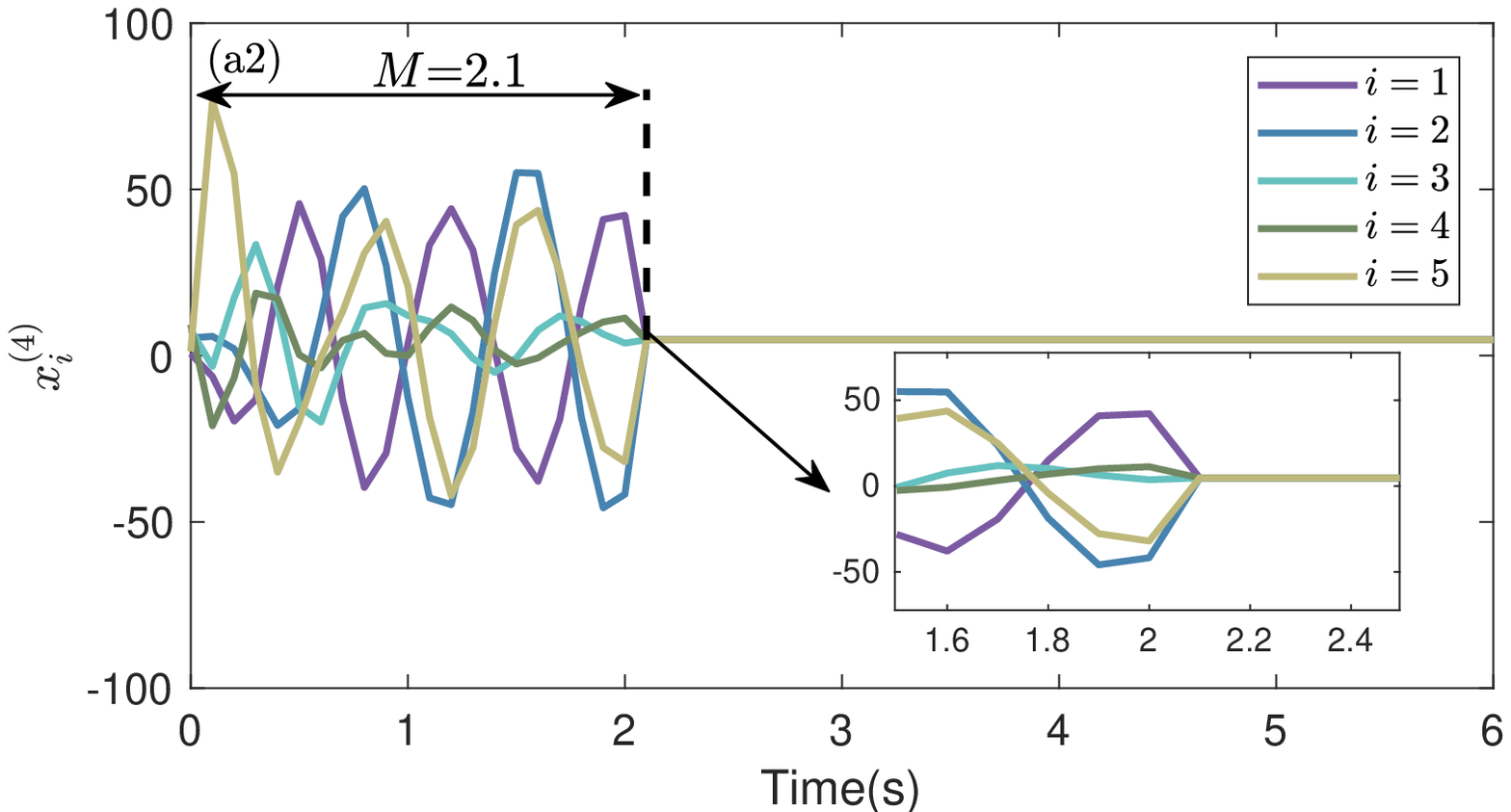} \label{figx4}
	\end{minipage}}
	\quad
	\subfigure[Temporal evolution of  $x_{i}^{(3)}$ for routine protocol (b1) and MDCP (b2).]{
		\begin{minipage}[b]{0.46\linewidth}
			\centering
			\noindent\includegraphics[width=0.9\textwidth]{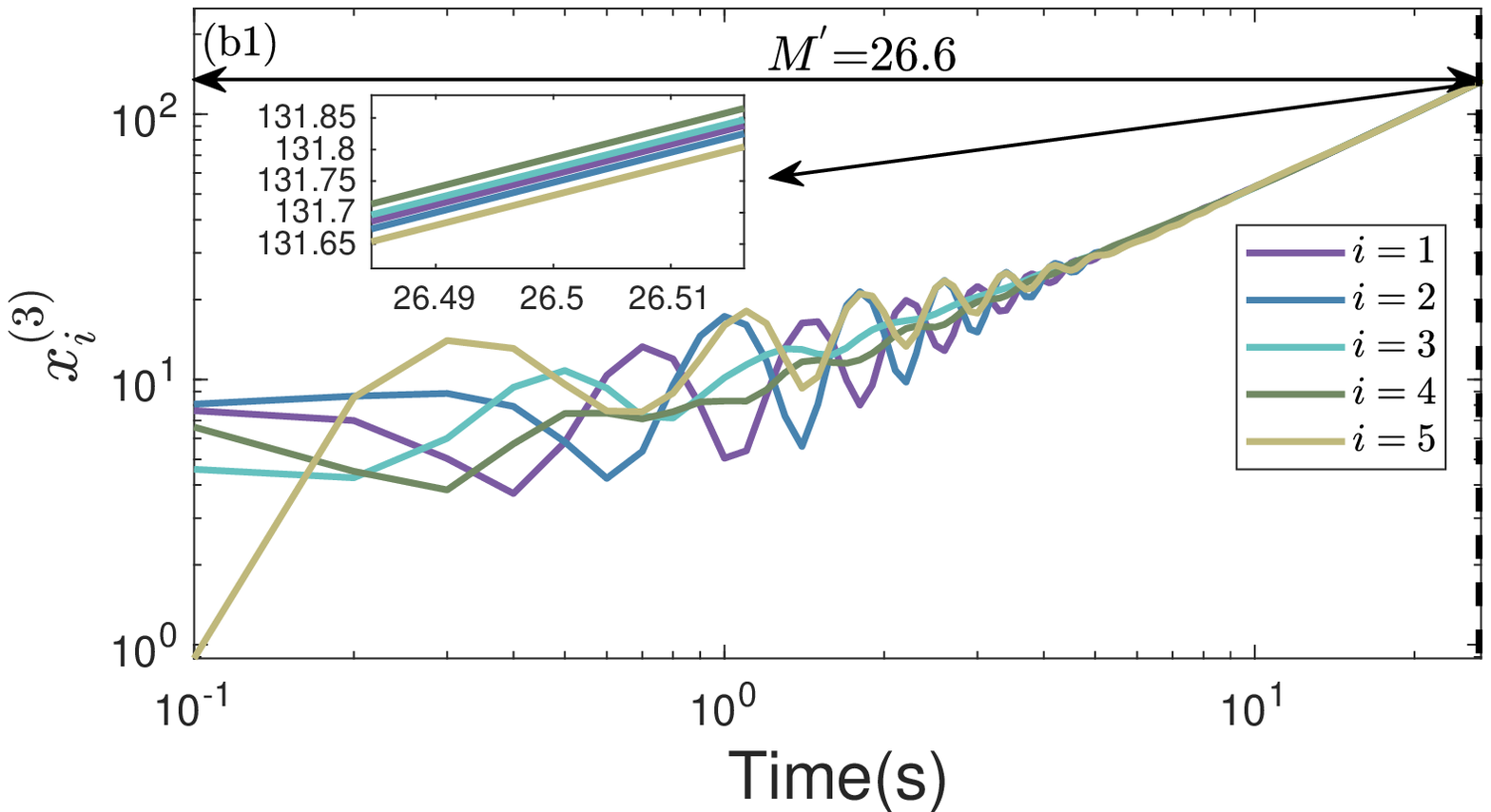}\
			\noindent\includegraphics[width=0.9\textwidth]{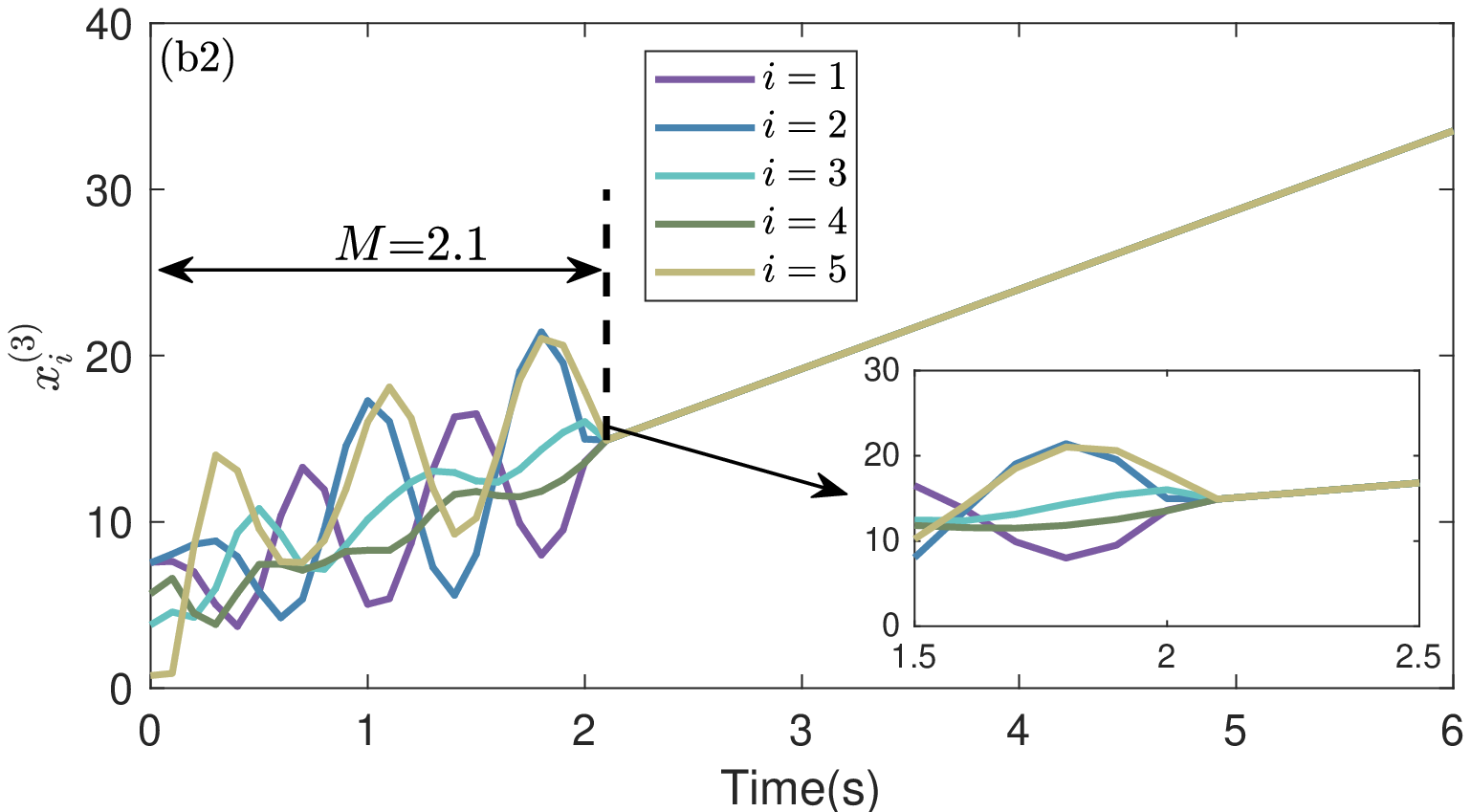} \label{figx3}
	\end{minipage}}
	\subfigure[Temporal evolution of  $x_{i}^{(2)}$ for routine protocol (c1) and MDCP (c2).]{
		\begin{minipage}[b]{0.46\linewidth}
			\centering
			\noindent\includegraphics[width=0.9\textwidth]{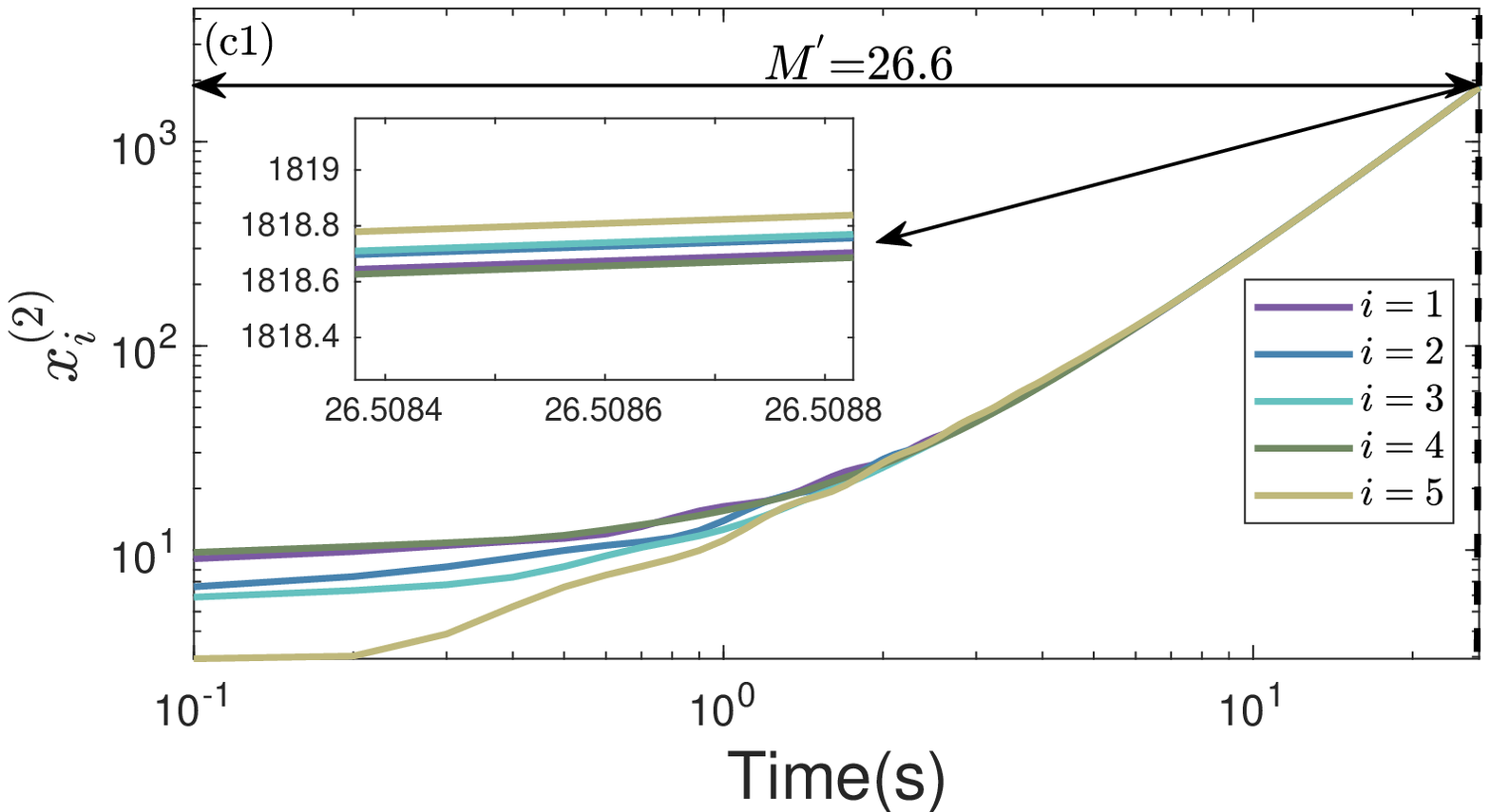}\
			\noindent\includegraphics[width=0.9\textwidth]{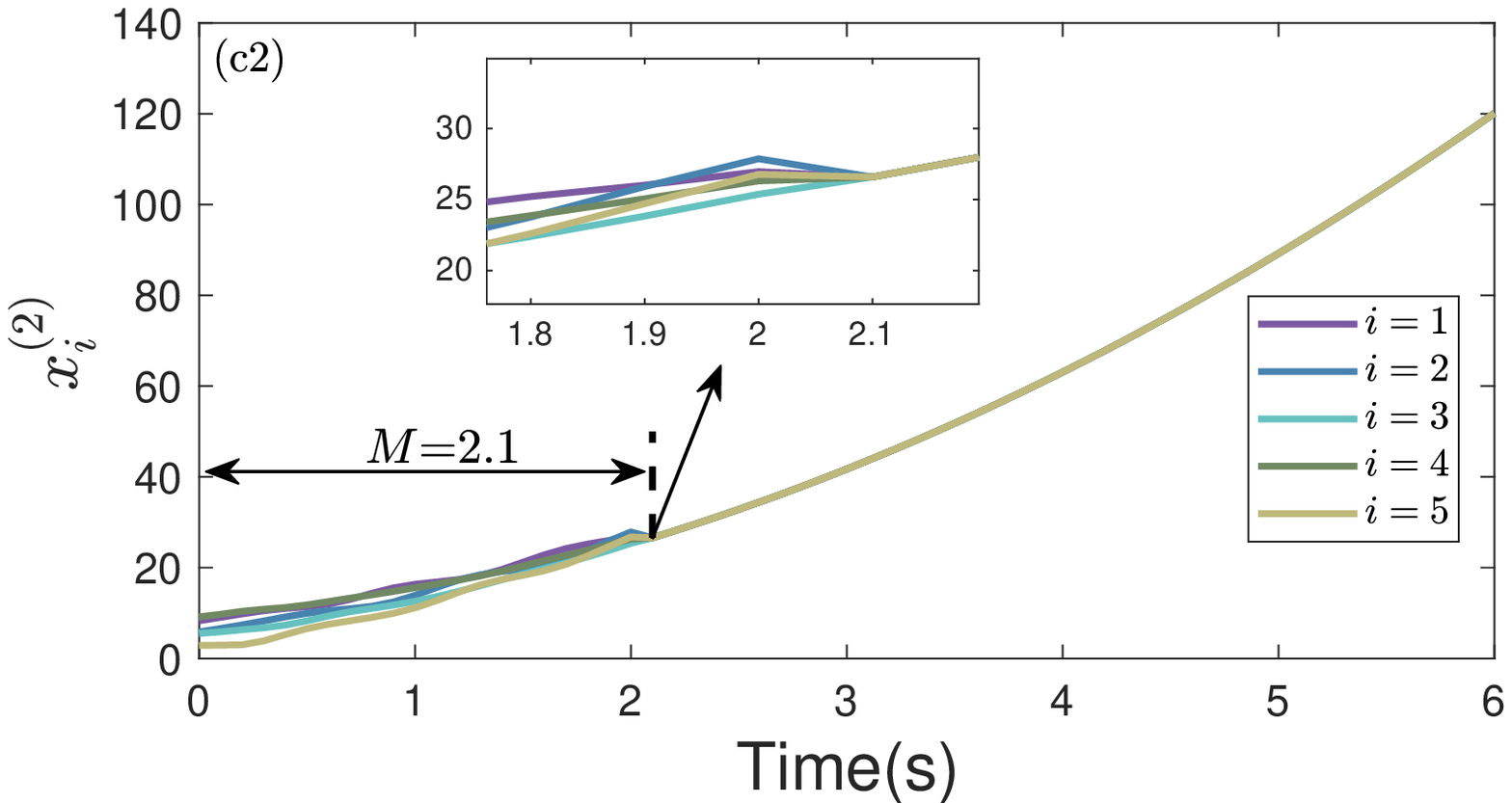}\label{figx2}
	\end{minipage}}
	\quad
	\subfigure[Temporal evolution of  $x_{i}^{(1)}$ for routine protocol (d1) and MDCP (d2).]{
		\begin{minipage}[b]{0.46\linewidth}
			\centering
			\noindent\includegraphics[width=0.9\textwidth]{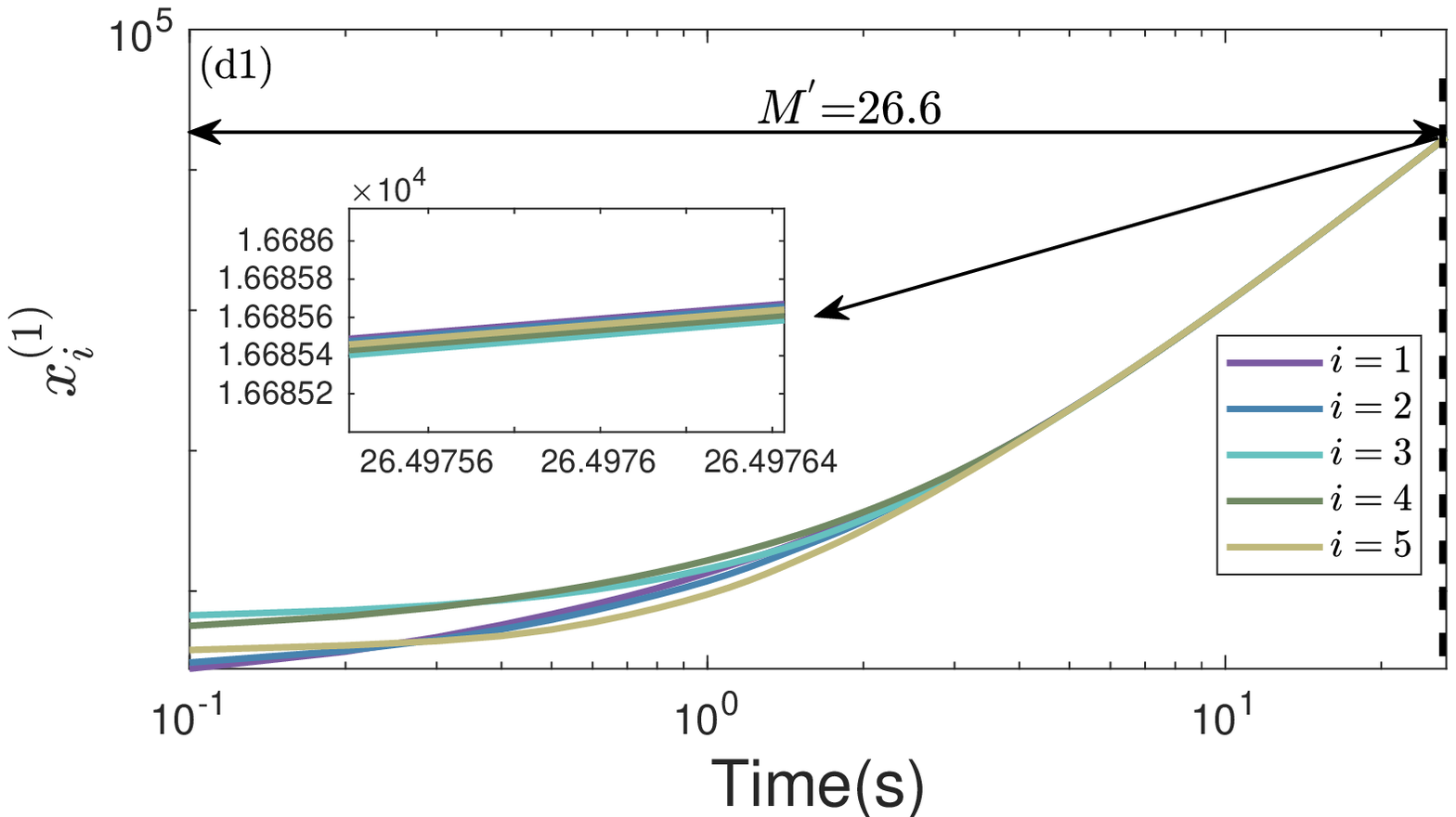}\
			\noindent\includegraphics[width=0.9\textwidth]{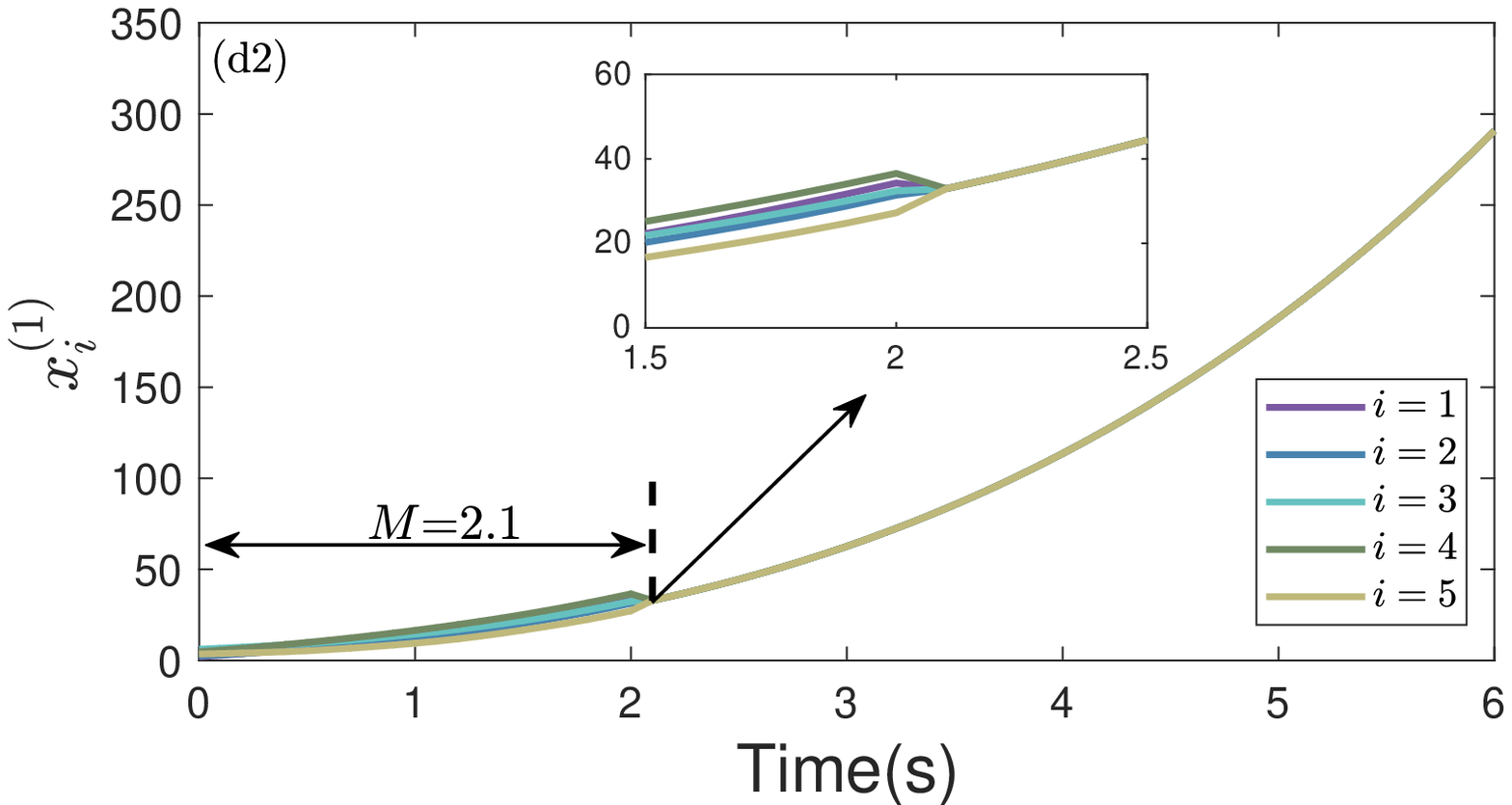}\label{figx1}
	\end{minipage}}
	\caption{ Temporal evolution of  state for the fourth-order MAS with topology shown in Fig.~\ref{topology}, the method in \citep{ren2007high} is adopted as the routine procedure. }\label{all4}
\end{figure*}

 \vspace{-0.5em} \hspace{0.2 in}Consider a discrete-time 4-order linear MAS  composed of five agents governed by (\ref{dynamic}) and (\ref{ui}), whose  communication topology is given in Fig.~\ref{topology}.	 Pick  the external coupling weight $\omega=-0.2$, internal   coupling weight $c_{0}=6,c_{1}=6,c_{2}=17,c_{3}=2$, sampling time $\epsilon=0.1$. The Laplacian matrix is given as follows
$$
\mathcal{L} = \left[\begin{array}{ccccc}
	2 &   0   & -1  &  -1   &     0\\
	-1 &  1    &     0  &       0  &  0\\
	-1 &        0  & 2   &     0 &   -1\\
	-1 &       0  &       0  & 2  &  -1\\
	0   & -1  &  -1  &  -1 & 3
\end{array}\right].
$$
 \vspace{-1em} \hspace{0.2 in}The initial values
\begin{align*}
	X(0)= [ & 1.9660  \quad
	2.5108 \quad
	6.1604 \quad
	4.7329 \quad
	3.5166  \\
	& 8.3083 \quad 
	5.8526 \quad 
	5.4972 \quad
	9.1719 \quad
	2.8584   \\
	& 7.5720 \quad 
	7.5373 \quad 
	3.8045 \quad
	5.6782 \quad
	0.7585  \\
	& 0.5395 \quad 
	5.3080 \quad
	7.7917 \quad
	9.3401 \quad
	1.2991  ]\t  
\end{align*}
are  generated randomly.

 \vspace{-1em} \hspace{0.2 in}According to (\ref{xk}), the corresponding Perron matrix is set as follows
$$
W=\left[\begin{array}{cccccc}
	I_{5} & 0.1 I_{5} & \mathbf{0}_{5} & \mathbf{0}_{5}\\
	\mathbf{0}_{5} & I_{5} & 0.1 I_{5} & \mathbf{0}_{5}\\
	\mathbf{0}_{5} & \mathbf{0}_{5}  & I_{5} & 0.1 I_{5} \\
	-1.2\mathcal{L}  & -1.2\mathcal{L} & -3.4\mathcal{L} & I_{5}-0.4\mathcal{L}
\end{array}\right].
$$

 \vspace{-1em} \hspace{0.2 in}Consider the first agent, i.e., $i=1$. Establish a Hankel matrix by (\ref{Gamma}), increase $\widehat{\overline{D}}_1$ until $\Gamma$ loses rank. Then, $\widehat{\overline{D}}_1=10$,  which implies  $2\widehat{\overline{D}}_1+1=21$ is the minimal individual memory length. Besides, $\Gamma^\bot$ is given as follows 
\begin{align*}
	\zeta_1= [ &0.0000,
	0.7209,
	-3.2848,
	7.5597,
	-10.8951,
	9.8706,\\
	&-4.0706,
	-2.6171,
	5.2873,
	-3.5668,
	1.0000 ]\t ,
\end{align*}
which is the coefficient of  (\ref{p_i}). Together with (\ref{qit}),  the minimal polynomial pair of $W$  is calculated below
\begin{align*}
	\alpha_{1}= [&0.0000 ,0.7209,-6.1685, 25.0245, -63.7266,\\  &112.6697,-142.4475, 124.0291, -59.0453, -14.2656, \\
	 &53.3886,-49.1670,25.5545, -7.5668,1.0000]\t .
\end{align*}

 \vspace{-1em} \hspace{0.2 in}Then, (\ref{Phi1}) can be obtained. According to (\ref{beta}), one has that
\begin{align*}
	[\beta_{0} ,\beta_{1} \dots \beta_{3}]\t= [  0.0048 \quad
	0.0538 \quad
	0.6762\quad
	3.7570
	]\t .
\end{align*}

 \vspace{-1em} \hspace{0.2 in}Besides, $b_{3-j} = \beta_j \text{ for }j =0,1,\cdots, 3$. According to (\ref{c}), the first-order consensus item of the first agent is computed as 
$ \widetilde{\wp}_c^{(1)}(k)=0.000794850061k^3 + 0.02451822315k^2 + 0.6508490022k + 3.757019522$. Here,  the result is converted from a fraction to decimal with 10 significant figures for convenience.  In addition, once the first agent has predicted the eventual consensus value, it  will send the  value to other  ones  simultaneously.

 \vspace{-1em} \hspace{0.2 in}To compare the performances of the present MDCP and the routine  consensus control protocol for high-order MASs \citep{ren2007high}  (in Abbr. Ren's protocol), let $M^{'}$, $M$ denote consensus-window-launch-time (CWLT) of Ren's  protocol and MDCP, respectively, here $\sigma=0.1$. Note that the smaller the $\sigma$ is the larger the CWLT is.  
 As shown in  Figs.~\ref{all4} (a2-d2),  deadbeat convergence to the consensus state is achieved  by the present MDCP.  By contrast, from  Figs.~\ref{all4} (a1-d1), it is observed that Ren's  protocol   yields asymptotic convergence to the eventual  same consensus value.
Thus, the main technical results of Theorem~1 and  2 are verified.

\begin{figure}
	\centering
	\includegraphics[width=0.42\textwidth]{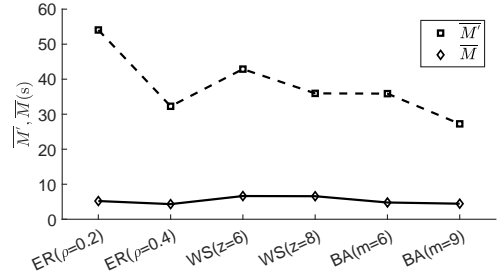}
	\caption{The average CWLT $\overline{M^{'}}$ and $\overline{M}$ computed for independent $\mathcal{S}=100$  $\mathrm{ER},\mathrm{BA},\mathrm{WS}$ networks with size $N=20$ independently.}
	\label{KM}
\end{figure}

 \vspace{-1em} \hspace{0.2 in}To examine the feasibility of the proposed MDCP on more general complex networks, we consider   benchmark 
Erdös–Rényi ($\mathrm{ER}$) \citep{ER}, Barabási–Albert ($\mathrm{BA}$) \citep{BA} and  Watts–Strogatz ($\mathrm{BA}$) \citep{WS}  networks. In  the $\mathrm{ER}$ network, node pairs are connected with a probability $\rho$. Initially, $\mathrm{BA}$ networks is a small clique of $m$ nodes, and at each time step, a new node is introduced and connected to $m$ existing nodes. $\mathrm{WS}$ network is a  lattice where each node connects to $z$ neighbors   with a rewiring probability  of $0.3$.
We generate  $\mathcal{S}=100$ networks of size $N=20$ for each  kind of  network model randomly  with randomly  initial individual state in the range $\left[0,30 \right] $. In each network, we  pick up an observed node $i$  from $20$  nodes  independent runs. Define the average  CWLT of $\mathcal{S}$  networks with $N$ agents as $\overline{M}=\frac{1}{N \times \mathcal{S}}  \sum_{j=1}^{\mathcal{S}}\sum_{i=1}^{N}M_{i, j} \text { and } \overline{M^{'}}=\frac{1}{\mathcal{S}} \sum_{j=1}^{\mathcal{S}} M^{'}_j$, respectively. Here, $\sigma=0.1$, $M_{i, j}$ denotes CWLT for the $i$-th agent of the $j$-th networks, and $M^{'}_j$ refers to CWLT of the $j$-th networks. It is observed from Fig.~\ref{KM} that the average  CWLT of MDCP is  much shorter than Ren's  protocol 
 ($\mathrm{ER}(\rho=0.2)$:   5.20s VS 54.05s,
  $\mathrm{ER}(\rho=0.4)$:   4.32s VS   32.26s,
  $\mathrm{WS}(z=6)$:  6.63s VS  42.88s,
  $\mathrm{WS}(z=8)$:   6.58s VS 35.94s,
  $\mathrm{BA}(m=6)$:  4.78s VS  35.87s,
  $\mathrm{BA}(m=9)$:  4.44s VS 27.24s),
which implies that consensus has been substantially  accelerated by locally sequential observations. Both the feasibility of Theorems~1 and 2, and the superiority of the present MDCP have thus been verified.

 \vspace{-0.5em}
\section{Conclusion}
 \vspace{-0.5em} \hspace{0.2 in}In this paper, both minimal-time deadbeat consensus and instant individual disagreement degree prediction problems are addressed by the propose MDCP for a class of  discrete-time high-order linear MASs with directed communication topological backbones. Deadbeat consensus is achieved to fulfill prompt coordination  requirements of modern industrial MASs. Sufficient conditions concerning the topology and the associate eigenspectrum of Perron matrix are derived to  guarantee the minimal-time deadbeat consensus and instant individual disagreement degree prediction. Extensive simulations are conducted on a variety of benchmark complex networks to show the effectiveness and superiority of the proposed MDCP protocol. The  present  MDCP can be expected to pave the way from minimal time  deadbeat consensus theory to prompt responses to emergence of natural/social/engineering swarming  systems.

\bibliographystyle{model5-names}
\bibliography{reference}\vspace{-1em}
 \vspace{-0.5em}
\begin{ack}                               
 \vspace{-0.5em} \hspace{0.2 in}This work is supported by the National Natural Science Foundation
of China under Grants 62225306 and U2141235.  
\end{ack}

\end{document}